\documentclass[preprint]{aastex}
\usepackage{graphicx}
\newcommand {\be} {\begin{equation}}
\newcommand {\ee} {\end{equation}}

\begin{document}
 
\title{Hardness-Intensity Correlations in Magnetar Afterglows}

\author{Feryal \"Ozel}

\affil{University of Arizona, Department of Physics, 1118 E. 4th St.,
Tucson, AZ 85721}

\author{Tolga G\"uver} 
 
\affil{Istanbul University, Astronomy \& Space Sciences Department,
Beyaz\i t, Istanbul, 34119}

\begin{abstract}
We explore the hardness-intensity correlations observed in several
AXPs and SGRs within the framework of a thermally emitting magnetar
model.  Using our detailed atmosphere models and taking into account
reprocessing of the surface emission by the magnetosphere, we show
that the hardness of the surface spectra increases with increasing
temperature and hence the changes in the effective temperatures of the
outer layers of the star alone can account for the observed
correlations. We conclude that the slow release of the heat deposited
in the deep crust during a magnetar burst naturally accounts for the
spectral changes during the afterglow. The correlations are further
enhanced by changes in the structures of the magnetic currents during
or following a burst. However, the additional hardening produced by
scattering of the surface photons off the magnetospheric charges
saturates at moderate values of the scattering optical depth.
\end{abstract}
 
\section{Introduction}

Anomalous X-ray Pulsars (AXPs) are a class of neutron stars identified
by a number of unique characteristics. Rapid secular spindowns
($\dot{P} \sim 10^{-13}-10^{-10}$~s~s$^{-1}$), clustered periods
ranging between $5-12$~s, the lack of observable companions, and
persistent X-ray luminosities that exceed their rotational energy
losses are among the known properties of these sources (see Kaspi 2006
for a review). Soft Gamma-ray Repeaters (SGRs), on the other hand, are
a related class of neutron star sources that are primarily defined by
recurrent, often highly super-Eddington bursts of soft gamma rays and
hard X-rays. Also detected as persistent X-ray pulsars, they share
many of the quiescent properties of AXPs (see Woods \& Thompson 2006
for a review).

The intense bursts of SGRs lend strong support to the identification
of these sources as ultramagnetic ($B \gtrsim 10^{14}$~G) neutron
stars, or magnetars (Duncan \& Thompson 1992; Kouvelioutou et al.\
1998). It is again this bursting activity that establishes the most
firm connection between AXPs and SGRs (Gavriil, Kaspi, \& Woods
2003). In AXPs, numerous bursts have been observed to date in about
half of the sources, with strengths ranging from strong outbursts to
long term flux variations (Kaspi 2006). Some of these bursts
accompanied other changes in the sources, such as increases in the
pulsed flux, large glitches, and spectral hardening.

Continuous monitoring of SGRs and AXPs following bursting activity has
revealed significant clues about the nature of the bursts and the
sources themselves. Initially, afterglow studies focused on the
variation of the source flux with time in order to measure the total
energy release and the cooling timescales (e.g., Kouveliotou et al.\
2003). These studies indicate that the location of the energy
deposition during bursting events is a few hundred meters beneath the
surface. Monitoring the post-flare pulse profiles has shown
significant changes that are likely to be related to magnetic field
reconfiguration (Woods et al.\ 2001). This is consistent with the
presence of glitches concurrent with the bursts, which also affect the
long-term spindown behavior of the sources (Kaspi 2006).

There are still a number of open questions regarding the underlying
mechanisms of the bursts and their long-term effects on the properties
of the magnetars. One such question is related to the correlation
between the hardness of the X-ray spectra and the flux variations that
have been recently observed in several sources (Mereghetti et al.\
2005; Rea et al.\ 2006; Tiengo et al.\ 2006; Campana et al.\ 2006;
Woods et al.\ 2006). The sudden increase and the subsequent slow decay
of the X-ray flux is understood to be the result of heat deposition in
the surface layers. The corresponding variations in the spectral index
have so far been attributed to the changes in the magnetic field
structure and the increase in the density of charges in the
magnetosphere, independent of the surface emission. This
interpretation has not taken into account that the surface emission
spectra have an intrinsic hardness which depends strongly on the
atmospheric temperature and thus can change as the magnetar cools.

In this Letter, we show that the changes in the effective temperatures
of the outer layers of the star alone can account for the observed
correlations. This is further enhanced by changes in the structures of
the magnetic currents during or following a burst. However, the
additional hardening produced by scattering off the magnetospheric
charges saturates at moderate values of the scattering optical depth.

\section{The Hardness of Magnetar Spectra}

In the following, we calculate the X-ray spectrum emerging from the
surface of a fully ionized, hydrogen atmosphere of a magnetar in
radiative equilibrium. We take the magnetic field to be locally
perpendicular to the neutron star surface. We take into account the
effects of vacuum polarization and proton cyclotron scattering as
discussed in \"Ozel (2003). The surface emission spectrum is
completely defined by the effective temperature $T_{\rm{eff}}$ of the
atmosphere, the surface magnetic field strength $B$, and the
gravitational acceleration $g$ on the stellar surface, which we set to
g=1.9$\times10^{14}$ cm$\,\rm{s}^{-2}$.

The surface radiation is reprocessed by the charges in the
magnetosphere of the star. The dominant effect is the scattering of
photons at radii at which they are resonant with the local electron
cyclotron energy (Thompson, Lyutikov, \& Kulkarni 2002; G\"uver,
\"Ozel, \& Lyutikov 2006). We calculate this effect using the Green's
function approach described in \citet{lg06}, assuming that the field
in the magnetosphere is spherically symmetric and follows a $1/r^{3}$
dependence. The emerging spectrum depends on two parameters: the
resonant scattering optical depth $\tau$ and the thermal electron
velocity $\beta$.

The atmospheric model spectra are broader than a blackbody and have
substantial structure in the continuum, even beyond the presence of a
proton cyclotron absorption line. In particular, at high photon
energies, the model spectra fall off more slowly than a blackbody,
which is typically referred to as hard excess. There are three
phenomena that determine the amount and hardness of this excess.  {\it
(i)} The frequency dependence of the free-free opacity in a magnetized
plasma causes the higher energy photons to originate deeper in the
atmosphere, where the temperature is higher. This dependence, however,
is weaker than that in a nonmagnetized atmosphere. If this were the
only mechanism that generates hard excesses, as is commonly believed,
magnetar spectra would be systematically softer than those of
nonmagnetic stars. {\em (ii)} The presence of a weak proton cyclotron
absorption line in the X-ray band significantly alters the continuum
spectrum in the nearby energies. As a result, the location of the
proton cyclotron line with respect to the peak of the spectrum affects
significantly its perceived hardness. This means that, even for a
fixed magnetic field strength, changing the effective temperature of
the atmosphere and thus whether the peak of the spectrum appears
above, at, or below the cyclotron line changes the hardness of the
spectrum. {\it (iii)} Vacuum polarization resonance produces a
depression in the spectrum, which, for the magnetic fields of
magnetars, occur at a few keV. Fitting such spectra with a blackbody
function leads to blackbody temperatures smaller than or comparable to
the effective temperature and significant hard excess.

The effects of all three phenomena discussed above depend strongly on
the temperature profile in the atmosphere, which is determined by the
effective temperature (i.e., the brightness of the magnetar when the
emitting area is constant). This is shown in Figure~1 for a magnetic
field strength of $8 \times 10^8$~G and for effective temperatures varying
between $0.3-0.6$~keV, which is in the range of temperatures inferred
for cooling magnetars. Note that while the rest of the calculations
presented here incorporate the effects of resonant scattering in the
magnetosphere, this figure shows only atmospheric spectra for
illustrative purposes. 

The effects of magnetospheric scattering on magnetar spectra have been
discussed in Lyutikov \& Gavriil (2006) and G\"uver et al.\
(2006). The non-Planckian tails already present in the surface spectra
become even harder due to resonant Compton scattering in the
magnetosphere and the equivalent widths of the cyclotron absorption
lines are reduced. These models then have all the qualitative features
required to fit most quiescent spectra of AXPs and SGRs.

\begin{figure}
\centering \includegraphics[scale=0.8]{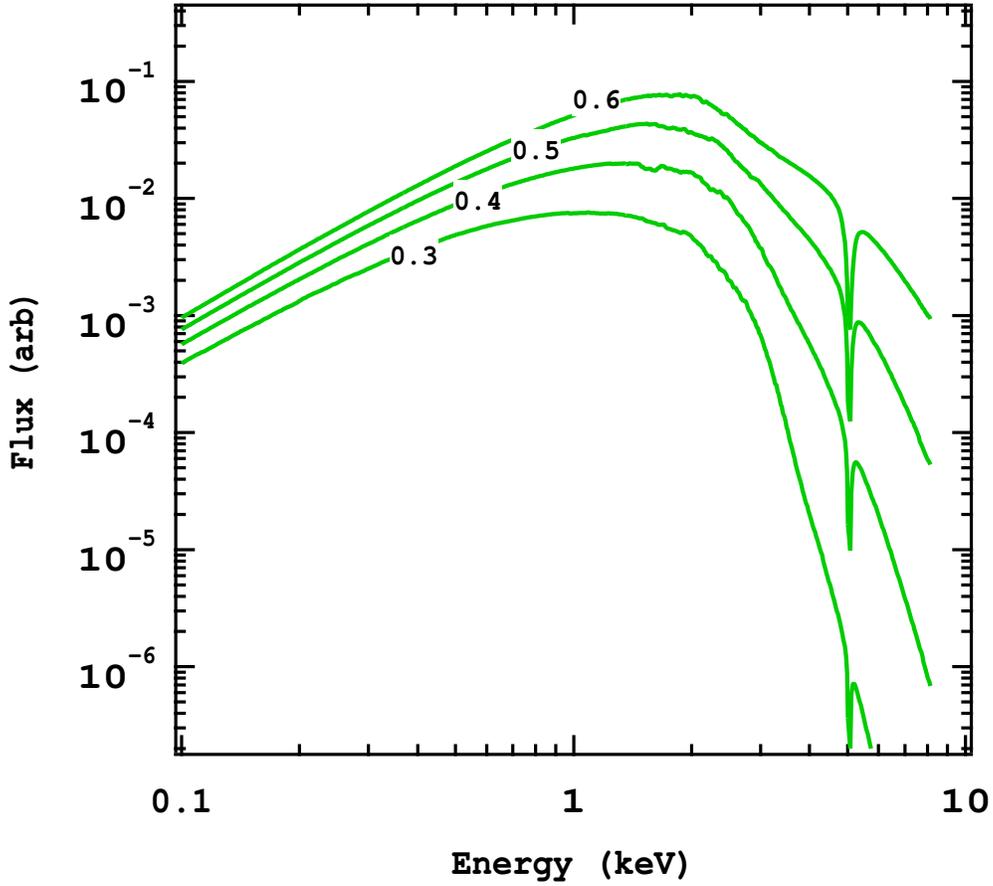}
   \caption{The spectra emerging from the atmosphere of a magnetar
with a surface magnetic field strength of $8 \times 10^{14}$~G, and
different values of the effective temperature in keV. This shows that
as the source flux increases, the spectra become harder. When the
surface emission is further reprocessed in the magnetosphere, the
spectra become harder and the equivalent widths of the cyclotron lines
are significantly reduced.}

\label{spectra}
\end{figure}

The hardness of the observed spectra of AXPs and SGRs has typically
been quantified in terms of the power-law index of a phenomenological
blackbody plus power-law fit (e.g., Rea et al. 2006; Woods \& Thompson
2006).  These studies yield power-law indices that range between $2-4$
for different sources and varying epochs. However, there are two
reasons why theoretical spectra cannot be meaningfully described in
terms of these two components. First, the power-law component
dominates over the blackbody at low photon energies, which is an
unphysical situation. This is particularly problematic for steep
power-law slopes often inferred from AXP and SGR spectra. In the case
of observations, this is prevented by taking into account the
interstellar extinction which attenuates the spectrum at low
energies. Similarly, theoretical spectra always cut off exponentially
beyond some characteristic energy scale while a phenomenological
power-law extends to infinite energies.

To quantify the hard excess theoretically, we instead define hardness
as the ratio of the flux in the hard band to the flux in the soft
band. We choose the fluxes in the $4-10$~keV and $2-4$~keV intervals
as the hard and soft bands, respectively, as hardness defined in this
way mimics the behavior of the power-law index closely for our
spectra. Note that the varying response of the detector at different
photon energies also has an effect on the hardness of the observed
spectra. We do not take into account the detector response in any of
the model predictions presented here, and therefore, would exercise
caution when performing direct comparisons with data.

\section{Spectral Manifestations of Crustal Heating}

The energy released in a magnetar burst is thought to have effects on
both the neutron star surface and its magnetosphere. The long
timescales observed in the afterglows suggest that most of the burst
energy is released several hundred meters deep in the crust
\citep{let02, ketal03}, altering the temperature profile of the
surface layers. Moreover, the burst can also reconfigure the
magnetosphere, change the density of the charged particles there, and
accelerate them to higher energies. All of these changes can affect
the source spectra following bursting activity.

Observationally, the spectra of SGRs during post-burst cooling and of
AXPs during different flux states show a correlated change between the
source brightness, the temperature of the phenomenological blackbody
component, and the hardness of the spectrum (Mereghetti et al.\ 2005;
Rea et al.\ 2006; Tiengo et al.\ 2006; Campana et al.\ 2006; Woods et
al.\ 2006). Specifically, as the brightness decreases, the inferred
blackbody temperature decreases and the spectrum becomes softer. This
indicates that independent of the changes that occur in the
magnetosphere, the temperature of the surface layers changes with the
changing flux of the magnetar. As we discussed in the previous
section, this alone alters the hardness of the spectrum emerging from
the atmosphere.

In order to disentangle the role of the temperature change in the
crust from the scattering optical depth change in the magnetosphere on
the hardness of the spectra, we focus on a setup in which only the
temperatures of the outer layers of the star increase, without any
changes in the magnetospheric configuration. We achieve this by fixing
the model parameters that describe the magnetosphere at values that
can qualitatively reproduce the quiescent spectra of AXPs and SGRs: we
take a scattering optical depth of $\tau=3$ and a thermal electron
velocity $\beta=0.3$ (G\"uver et al.\ 2006). (We will present the
quantitative results obtained from fitting observed spectra with these
models in a forthcoming paper.) We model the flux enhancement as a
change in the effective temperature $T_{\rm eff}$ of the atmosphere
because the burst energy is deposited well below the photosphere.  We
vary $T_{\rm eff}$ over a wide range ($0.2-0.6$~keV) in our
calculations to model the large flux variations observed in some
afterglows (Kouveliotou et al.\ 2003). However, we do not exceed
temperatures above which noncoherent scattering {\em within} the
atmosphere becomes important or reach low enough temperatures at which
the assumption of complete ionization is not valid.

\begin{figure}
\centering \includegraphics[scale=0.8]{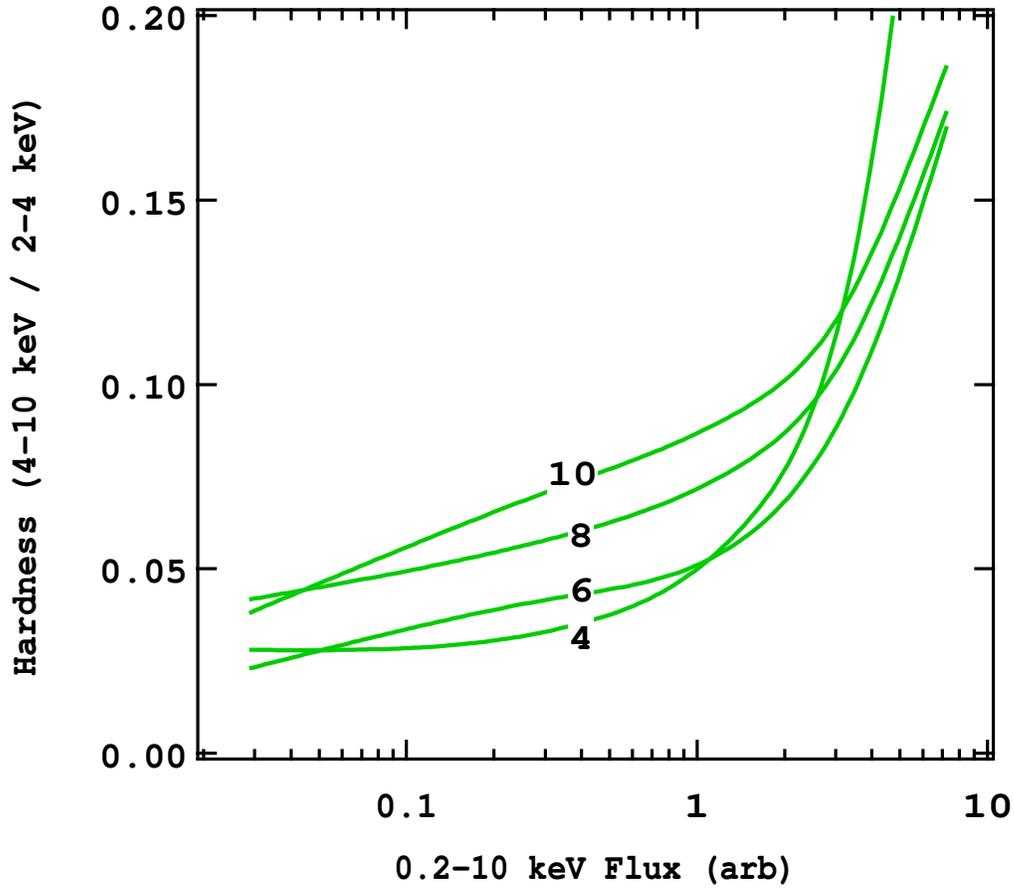}
   \caption{The hardness of the spectra emerging from the atmosphere
of a magnetar as a function of the $0.2-10$~keV source flux, where
hardness is defined as the ratio of the flux in the $4-10$~keV band to
the flux in the $2-4$~keV band. Different curves correspond to
different surface magnetic field strengths in units of $10^{14}$~G.
The spectral hardness increases with increasing flux, as in the
correlations observed for several AXPs and SGRs. }

\label{hr}
\end{figure}

The resulting correlated change in the hardness of the spectrum and
the $0.2-10$~keV flux is shown in Figure~\ref{hr}. For any magnetic
field strength, the hardness of the spectrum changes considerably even
with a modest change in the effective temperature. Furthermore, the
models follow the same trend as the observed spectra, i.e., they
become softer as the source flux decreases. The weak dependence of the
hardness on the magnetic field is a result of the various atmospheric
and magnetospheric effects discussed in the previous section.  

\section{Discussion}

In this paper, we show that the hardness of the theoretical spectra
emerging from the surface of a magnetar is a strong function of its
luminosity, as characterized by the effective temperature of the
atmosphere. This provides a natural explanation of the observed
correlation between the spectral hardness and X-ray flux found in a
number of AXPs and SGRs (Mereghetti et al.\ 2005; Rea et al.\ 2006;
Tiengo et al.\ 2006; Campana et al.\ 2006; Woods et al.\ 2006).

\begin{figure}
\centering \includegraphics[scale=0.8]{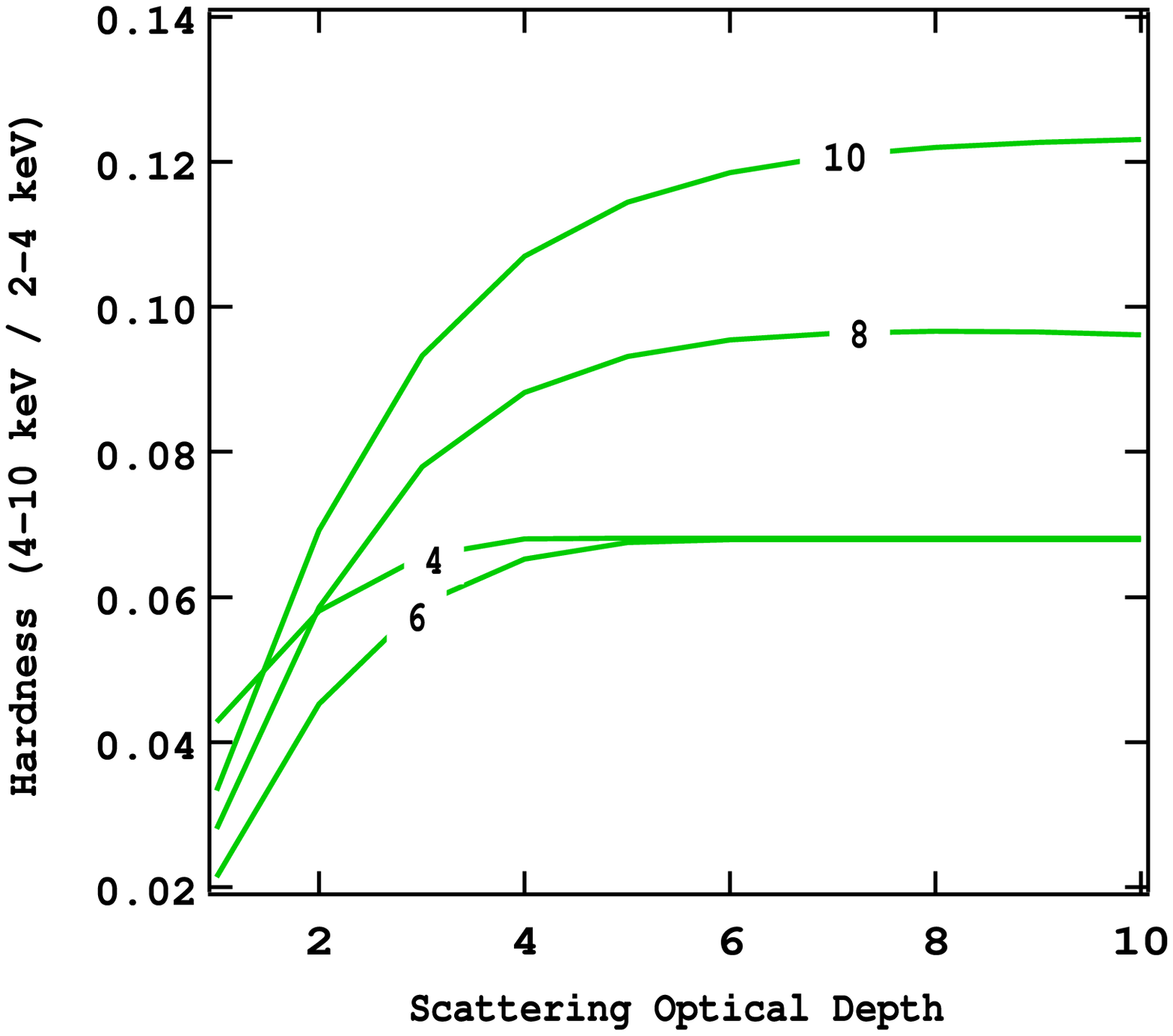}
   \caption{The dependence of the hardness of the magnetar spectra on
the scattering optical depth in the magnetosphere for different values
of the magnetic field strength in units of $10^{14}$~G. The hardness,
which is defined in Figure~2, saturates at moderate values of the
optical depth. }
\label{hrtau}
\end{figure}

The changes in the charged particle density and energetics in the
magnetosphere of a magnetar during a burst can also alter the spectral
hardness. This is shown in Figure~\ref{hrtau} where the spectral
hardness is plotted as a function of the scattering optical depth in
the magnetosphere. Contrary to the case of varying the effective
temperature of the atmosphere, increasing the scattering optical depth
produces a more modest change in the spectral hardness. Moreover, this
effect saturates at optical depths $\tau \gtrsim 5$, as is often the
case for Compton upscattering studied in different astrophysical
settings (Sunyaev \& Titarchuk 1980). This is a consequence of our
assumption of a nearly constant particle velocity in the
magnetosphere. Assuming a multidimensional magnetic field structure
and a broad spectrum of particle velocities can further change the
shape and hardness of these spectra. Fernandez \& Thompson (2006)
carried out detailed Monte Carlo simulations of magnetospheric
scattering and explored these possibilities.

Our results demonstrate that atmospheres play a significant role in
determining the hardness of magnetar spectra and its correlated change
with source flux. Therefore, when performing detailed simulations of
magnetar spectra or interpreting observational results, the
non-Planckian shape of atmospheric spectra need to be
considered. Furthermore, the complicated beaming patterns of the
radiation emerging from the atmosphere as well as general relativistic
effects (\"Ozel 2002) can be used in conjuction with the
magnetospheric models to also predict the evolution of pulse profiles
with bursting activity. Comparison of such detailed models with the
observations of the spectra, pulse profiles, and cooling timescales of
magnetar burst afterglows offers the possibility of disentangling the
effects of the burst on the crust and the magnetosphere.

\end{document}